\documentclass{ws-ijmpa}
\usepackage[super,compress]{cite}
\usepackage{graphicx}
\usepackage{hyperref} 
\usepackage{CJKutf8} 
\usepackage[utf8]{inputenc}
\usepackage[T1]{fontenc}
\def\draftnote{}
\begin{document}
\markboth{Y. Shi}{From quantum entanglement to parity nonconservation}
%
%

\title{Scientific Spirit of Chien-Shiung Wu: \\From Quantum Entanglement to Parity Nonconservation
}

\author{Yu Shi}

\address{Wilczek Quantum Center, Shanghai Institute for Advanced Studies, Shanghai 201315, China \\
University of Science and Technology of China, Hefei 230026, China\\
Department of Physics, Fudan University, Shanghai 200433, China\\yushi@fudan.edu.cn }

\maketitle


\begin{abstract}
In 1950, Chien-Shiung Wu and her student published a coincidence experiment on entangled photon pairs 
that were created in  electron-positron annihilation. This experiment precisely verified the prediction of quantum electrodynamics. Additionally, it was also the first instance of a precisely controlled quantum entangled state of spatially separated particles,  although Wu did not know about this at the time.  In 1956,   Wu initiated  and led the so-called Wu experiment, which discovered parity nonconservation, becoming one of the greatest experiments of the 20th century. As Chen Ning Yang said, Wu's experiments were well known for their precision and accuracy. Experimental precision and accuracy manifested Wu's scientific spirit, which we  investigate here in some detail.  This paper is the translated  transcript of the speech the  author made at  the  International Symposium Commemorating 
the 110th Anniversary of the Birth of Chien-Shiung Wu, on May 31, 2022. The above abstract is the translation of the original abstract of  the speech.
 \end{abstract}
 

\ccode{PACS Nos: 03.67.Mn, 12.20.Fv, 01.65.+g}  


\vspace{1cm}

This paper is the English written version of a speech at the International Symposium Commemorating the 110th Birth Anniversary of Chien-Shiung Wu,  held on May 31, 2022 and organized by Southeast University (one of the descendents of the alma mater of Wu), in which speeches were given by  Chen Ning Yang, Tsung-Dao Lee, Shing-Tung Yau,  Wu's son Vincent Yuan,  along with many  distinguished physicists and  scholars~\footnote{A poster of this  symposium and the pdf file of the slides of the present speech, with text translated to English, are  ancillary files available on the abstract page of the present paper at {\em arxiv.org}.  The program and the abstracts of speeches of the symposium  are  available at  \href{https://mp.weixin.qq.com/s/9l4FTpHErzS8IED-zqs6hQ}
{https://mp.weixin.qq.com/s/9l4FTpHErzS8IED-zqs6hQ}. 
  The video of the whole conference is available at \href{https://news.sciencenet.cn/htmlnews/2022/5/479979.shtm}
{https://news.sciencenet.cn/htmlnews/2022/5/479979.shtm}. 
The video of the present speech is also available at \href{https://www.koushare.com/video/details/164531}
{https://www.koushare.com/video/details/164531}.  The  Chinese slides of the present  speech are available at  \href{https://mp.weixin.qq.com/s/lilZY3ClxvfP9Q8DSt2j_A}
{https://mp.weixin.qq.com/s/lilZY3ClxvfP9Q8DSt2j\_A}.
The Chinese transcript was published in {\em Micius Forum} on June 2, 2023,  available at \href{https://mp.weixin.qq.com/s/inwKeaTyI9FxT-qEEgsKDA}
{https://mp.weixin.qq.com/s/inwKeaTyI9FxT-qEEgsKDA}.  
 }.

We review the scientific contributions and scientific spirit of Chien-Shiung Wu. In particular, we emphasize the two contributions indicated in the title, i.e., quantum entanglement and parity nonconservation.

\section{Overview of Chien-Shiung Wu's academic career and research} 

First,  I give a comprehensive overview of all the research that Chien-Shiung Wu performed throughout her academic career.  For this purpose,  I looked  at all her papers with the help of the Web of Science, so the references listed in the slides are in the format of this database.

\subsection{China}

Chien-Shiung Wu graduated from the National Central University in 1934. Her  thesis was entitled ``Verification of the Bragg diffraction equation of  X-rays in crystals''~\cite{1}, and was supervised by Shih-Yuan Sze.  After working  for a year as a teaching assistant in the Department of Physics at Chekiang University,  with the recommendation of  the department head, T. C. Chang, she joined the Institute of Physics, Academia Sinica, in Shanghai, where she studied gas spectroscopy under the supervision of  Z. W. Ku~\cite{2}.

The fields of physics Chien-Shiung Wu was initially exposed to through her  undergraduate thesis work and one-year work in Shanghai  were  not  in nuclear physics,  but had much in common with radioactivity. Shih-Yuan Sze, who was only three years older than Chien-Shiung Wu, had worked on nuclear spectroscopy under the supervision of   Madame Curie, earning a Doctor of Science. This was also in the same area as Wu's later research on $\beta$ rays and nuclear physics. Z. W. Ku  was the first female Doctor of Physics in China and studied spectroscopy under the guidance of D. M. Dennison at the University of Michigan.

As a result of Z. W. Ku's recommendation and advice, Wu was admitted  by the University of Michigan to study at her own expense, and was financially supported by her uncle. On her way to Michigan, Wu visited Berkeley, where she was so impressed, especially by Ernest O. Lawrence's cyclotron, that she  wanted to stay in Berkeley. The cyclotron had been invented by Lawrence, so  it was an ideal place for studying physics. Another important factor that influenced Wu's decision was that she cared a lot about gender equality, and there was gender discrimination at the University of Michigan. In addition,  there were a lot of Chinese students at the University of Michigan at the time,  and Wu  didn't want her socializing be dominated  by  fellow Chinese students. So she stayed in Berkeley.  Her decision reflected her devotion to physics as a woman. 

\subsection{Berkeley} 

Let us now look at Wu's  PhD and postdoctoral work at Berkeley. Her PhD supervisor was Ernest O. Lawrence, and Wu's first published work was on $\beta$ decay under Lawrence's guidance~\cite{3}. The process $\beta$ decay is the decay, governed by the weak interaction, of neutrons into protons,  antineutrinos and electrons, with the latter called $\beta$ rays.  Wu studied X-rays excited by $\beta$ rays and emitted by radioactive lead. Interestingly, this part of her PhD linked her previous undergraduate thesis work and the main area of her career, $\beta$ decay, including her future  experiment on parity nonconservation.

Wu proceeded to do research with Emilio Segr\`{e}  for a few years, obtaining her PhD in 1940, and then she stayed on  with Segr\`{e} for a further two years as a postdoc.  She was Segr\`{e}'s first student after he moved to the United States from Italy. So while her nominal PhD supervisor was Lawrence, Segrè was her mentor. 

One of the very important pieces of work of Wu under Segr\`{e}'s supervision, which was another part of her PhD thesis work, was to study the radioactivity of xenon. This was produced in the fission of uranium, by using neutrons produced in cyclotrons, and involved identifying  two radioactive isotopes of xenon.  This work was published as two papers in 1940~\cite{4,5}. Some data from that work was unpublished at that time because of confidentiality considerations.

Later, this data  played a large role in the Manhattan Project,  which was the programme to develop the world's first atomic bomb during World War II. It was discovered,  that in nuclear reactors, the chain reaction producing plutonium from uranium and xenon was hindered or even stopped soon after it started, and was commonly known as ``poisoning''. Enrico Fermi and others found that this was because the xenon absorbed neutrons, which were needed in large quantities for the chain reaction to continue. So the  data on the absorption of neutrons by xenon isotopes were needed, which Segr\`{e} disclosed had been obtained by the pre-war work of Wu and himself.  Plutonium was the raw material for the plutonium bomb that was later dropped in Hiroshima, and Wu's data played a big role 
in this. In 1945, Wu and Segr\`{e} used the results of the pre-war experiments to publish a paper~\cite{6}.

During her postdoctoral period, Wu also used a cyclotron to produce neutrons for the study of radioactive mercury, in collaboration with Gerhart Friedlander, a graduate student of Professor (and future Nobel Prize winner) Glenn T. Seaborg~\cite{7,8}.

\subsection{1942 - 1948 and Wartime work} 

After her postdoc work was completed  Wu went, in 1942, to the eastern United States and did two years'  teaching. The first year was at Smith College (a women's college),  where Nancy Reagan was studying theatre at that time. Also at 
Smith was Xide Xie (later  the President of Fudan University, in Shanghai),   who came five years later to do her master's degree under the tutelage of Gladys Anslow,  who had known Wu at Berkeley and  hired Wu at  Smith~\cite{2}. Wu carried out her second year of teaching   at Princeton University,  where she taught a group of military officers for a few months. Thanks to the influence of Lawrence, who won the Nobel Prize in 1939 and played an important role in Manhattan Project, Wu joined Columbia University in 1944 working on the  Manhattan Project and she stayed on after the war.

Wu worked in the  department headed by John R. Dunning, who was an important  figure in the Manhattan Project and was responsible for gaseous diffusion separating uranium isotopes. Natural uranium is a mixture of two isotopes, 235 and 238, and to build a uranium bomb, it must be enriched with 235. However, the uranium bomb later dropped on Nagasaki was made from electromagnetically separated uranium 235 at the Oak Ridge Laboratory, rather than gaseous diffusion separation.  Later, two students of Dunning, James Rainwater and William W. Havens Jr., went on to become professors at Columbia University before Wu, and Rainwater received a Nobel Prize for his work on the theory of nuclear structure. These three people were collaborators of Wu. For many years after the war,  they were acknowledged in Wu's papers not coauthored by them. . 

They contributed a lot to the Manhattan Project, but because the wartime work was classified,  published a series of papers only  after the war.  From 1946 to 1948, they published about 8 papers on the slow neutron effect of hydrogen and other elements. From 1947 to 1948, they published about 7 papers  on  sensitive radioactive detectors of $\beta$ rays,  photons and other particles.  Wu's experience on this research had a very deep influence on her later career. So in addition to her  pre-war data on xenon isotopes, Wu had a great deal of other work to contribute directly to   the Manhattan Project.

These were essentially work done during the war, with papers published one after the other after the war, some with extensions.  This work has been often misunderstood as having started after the war.

\subsection{Wu's $\beta$ Decay Research Before the Parity  Revolution} 

After the war, Wu also began to open up a new field, that of $\beta$ decay. She studied the physics of $\beta$ decay, and compared her experiments with theory. Even before discovering parity nonconservation in weak interactions, Wu had already  published about 53 papers on $\beta$ decay or closely related subjects, such as the capture  of electrons by the atomic nucleus.

Wu  became a leading expert on $\beta$ decay, and published more than 50 papers on this subject decay during this period. Tsung-Dao Lee  once said that in the field of $\beta$ decay, no one could compare with Chien-Shiung Wu~\cite{9}. However,  she only became an associate professor in 1952~\cite{2} and a full professor in 1958~\cite{9}  after the discovery of  parity  nonconservation,  probably  because of  unfair treatment she received as a woman.

During this period, Wu's most important work was the verification of Fermi's theory of $\beta$ decay~\cite{10}. This is one of the great theories in the history of physics and was Fermi's greatest theoretical contribution, extending quantum electrodynamics to weak interactions at lower energies. Starting with the Fermi's  $\beta$-decay theory, quantum field theory was transformed from  a formal theory initially proposed by Wigner and Pauli  to a physical theory, making  the creation and annihilation of particles physical concepts, and ultimately a theoretical framework for particle physics. At that time there were also competing theories, such as those of Konopinski and Uhlenbeck.  Wu found that previous validation experiments had been done poorly, with samples  too thick. In her own experiment with R. D. Albert, in which cobalt 60  ($^{60}Co$) was attached to a thin layer, they was very careful and precise. She was experienced in making instruments and  skilled in  assembling  and operating them, and these played  a great role in her work. Indeed, these strengths were very important factors also in her later success.

In the words of  Chen Ning Yang, ``Fermi's theory was proved correct! Wu became the leading scientist in the field of $\beta$ decay experiments!"~\cite{11}. One can see the meticulousness of Wu's experiments, and her skills of assembling and operating the instruments.

\subsection{Positronium, electron-position annihilation and quantum entanglement}

Prior to the discovery of parity violation,   a relatively small amount of Wu's work   was on electron-positron annihilation. An electron and a positron  can form positronium, which is similar to an atom-like structure, and  which annihilates  when the electron and the positron come together. From 1950 to 1957, Wu published 6 papers on this subject. 

The first work of Wu on  this subject was to accurately verify the predictions of quantum electrodynamics with her student Irving Shaknov~\cite{12}. This paper was submitted in the same year as the one testing Fermi theory, and published the following year. 

In retrospect, this work is closely related to a currently active area, namely the realisation of quantum entanglement of photon polarization. The purpose of  Wu and Shaknov was  to test  the predictions of quantum electrodynamics, however, it also became a pioneering work on quantum entanglement. Later I will review more about this work.

\subsection{Parity nonconservation} 

Certainly, the most important work of Wu was the experiment on parity nonconservation, that was carried out with  her collaborators in 1956 and early 1957, and is known as the ``Wu experiment''. In 1956,   Chen Ning Yang and Tsung-Dao Lee   proposed the possibility of parity nonconservation in  weak interactions, and suggested a few experiments~\cite{13}.  Thanks to their discussions with Wu and Goldhaber,  one of the suggested experiments was the $\beta$ decay of polarized cobalt 60 nuclei. 
In collaboration with several low-temperature physicists at the U.S. Bureau of Standards,  Wu completed the experiment and found that parity was \textit{not conserved}~\cite{14}, which was a revolutionary breakthrough. Being polarized means that the spin of the nuclei has a definite value. Whether or not parity is conserved was shown here by whether or not the number of electrons decaying out of the nuclei, along and against the direction of the spin of the nuclei, are equal. The Wu experiment found that they were not equal.

We note that $\beta$ decay of  cobalt 60 had already been studied by Wu in testing Fermi theory~\cite{10}.

\subsection{Wu's  research  after parity revolution} 

Chien-Shiung Wu kept up with the times, and did a lot of  work  after the parity  revolution.

In 1957,  Wu published a paper on gas scintillation counters for detecting slow neutrons.

Wu continued to work on $\beta$ decay for many more years. From 1962 to the 1985, she published about 16 papers  on $\beta$ decay, including pioneering work on double $\beta$ decay. The subject of double $\beta$ decay is still a frontier in particle physics today, because if there are no neutrinos in double $\beta$ decay, the neutrinos are Majorana fermions, which are the antiparticles of themselves, so that the neutrinos produced by one $\beta$ decay are immediately absorbed by the other $\beta$ decay. This problem  is so far inconclusive, and experiments are still being done. 

The most important work of  Wu in $\beta$ decay during this period was on the so-called vector current  conservation, which she and her students confirmed in 1963. 

In the 1970s Wu did a lot of work with the M\"{o}ssbauer effect to study haemoglobin or other substances. From 1973 to 1978, she published about 15 papers on the subject.  In this way, Wu was also a pioneer in the field of biological physics.

In the 1970s and 1980s, Wu also worked on so-called exotic atoms, in which  an electron  is replaced by another particle, such as muon, pion or another meson. People  studied the  interaction between this particle and the nucleus. From 1974 to 1984, Wu published about 23 papers  on  exotic atoms and muon capture of nuclei. 

In the 1970s,  Wu returned to electron-positron  annihilation. It was noted that  her previous work with Shaknov showed the existence of quantum entanglement. After the Bell inequality was proposed,   Wu and students tried extending  their  previous work into one that tested Bell's inequality, and published a paper in 1975.  We will review it below. 

Wu also worked on techniques of nuclear orientation. In the 1957 experiment she led  on parity nonconservation, where it  was crucial to  polarize the nucleus. This is also called nuclear orientation, i.e., to orientate the spin of the nucleus, which was achieved by cryogenic techniques. Later on,  Wu specialized in this technique and its application in nuclear physics, especially for the study of nuclear structure. From 1976 to 1985, She published about six papers on this subject. 

\subsection{ Conservation of vector current}

In 1956 Yang and Lee  raised the possibility that the parity may not be  conserved in weak interactions, and then the experiment led by Wu found that parity  is indeed not conserved. So what exactly does a weak interaction theory look like? How to develop  a theory of weak interactions? How to extend and improve Fermi's $\beta$ decay theory into a general weak interaction theory? Three groups of scientists at the time proposed vector-axial vector theories, also called universal Fermi interaction theories. One of these groups was Feynman and Gell-Mann, whose theory had a key assumption called conservation of vector current, which explained that the coupling constants for different weak decays, e.g.  $\beta$ and $\mu$ decays,  were equal. This is because the protons and neutrons in the nucleus  have to be renormalized,  so some imaginary particles are created, which also have to take part in the weak interaction. The consequence is that the vector current is conserved, resulting in the same coupling constants for $\beta$ and muon decays. 

Gell-Mann pointed out that the theoretical assumption of conservation of vector current leads to the so-called weak magnetism.  
This is similar to the difference in magnetic moments of protons and neutrons in quantum electrodynamics, and can be verified by direct experimental measurements using decays of  $^{12}B$ and $^{12}N$.  But the experiments  had been  unsuccessful. 

Enter Wu in 1963. Wu and her students Y.K. Lee and L.W. Mo completed such an experiment~\cite{15}, which confirmed weak magnetism, thus verifying the conservation of vector current and contributing to the establishment of the electroweak unification theory.

Wu once concluded in a review article: ``The conserved  vector current hypothesis  for weak interactions has been confirmed experimentally and has served as a  guiding principle for the modern gauge theory of strong and weak interactions, where it is predicted as a  necessary consequence of vector gauge  theory.''~\cite{16}  Wu was at the forefront of the subject as an experimental physicist.

In a retrospect at 2015, Yang listed four papers of Chien-Shiung Wu,  the one verifying Fermi theory, the one on photon correlation or quantum entanglement,  the one on parity violation, and the one on conserved vector current~\cite{11}.    

\subsection{President of the American Physical Society } 

Chien-Shiung Wu served as the President of the American Physical Society from 1975 to 1976. One of her articles is her  speech as President in 1976~\cite{17}, from which  we can learn what American physicists were doing at that time. Nuclear Physics: understanding the charge distribution of atomic nuclei through electron scattering; scattering of protons by atomic nuclei; atomic nuclei absorbing pions and jumping to rotating highly excited states; $\mu$ atoms (an exotic  atom with a muon replacing an electron in the atom) verifying quantum electrodynamics; parity  nonconservation in nucleon scattering; heavy ion collisions to study rapidly rotating atomic nuclei, nuclear disintegration. Cosmic rays: possible magnetic monopoles. Synchrotron radiation applications in biology, solid state physics, chemistry. Biophysics: addition of heavy atoms to protein molecules, X-ray diffraction; with respect to haemoglobin, both synchrotron radiation and the M\"{o}ssbauer  spectra from Columbia group  revealed  differences in various cases. Solid state physics: measuring local magnetic fields in solids with muons. Particle physics: $g-2$ experiments with muons to test lepton universality; positronium; charmonium $J/\psi$.

Among the areas mentioned above, Wu's own group was involved in the study of exotic atoms; the magnetic monopole has not  been found to the present day; the study of haemoglobin by M\"{o}ssbauer spectroscopy was a result of  Wu's own group; the $g-2$ experiment of muons to test lepton universality is an important area that has continued to this day, with some most recent advances; the charmonium  (consisting of a charm quark and an anti-charm quark) was  discovered by Samuel C. C. Ting's   group and the Richter group in 1974, with the names $J$ and $\psi$ respectively,  and later called  $J/\psi$, with Ting and Richter sharing the 1976 Nobel Prize in Physics.

\section{Details of Wu's work on electron-positron annihilation and quantum entanglement} 

We now look at the details of Wu's early work on electron-positron annihilation and quantum entanglement.

In 1946, in order to test quantum electrodynamics, John Wheeler suggested to study the annihilation of positroniums, each consisting  of a positron and an electron,   and to probe the photon pairs created.  The annihilation comes mainly from the spin singlet state of the positronium, i.e., the quantum state with total spin 0. Therefore, if the orbital angular momentum is also 0, then the total angular momentum is 0,   thus for the two photons produced from  the positronium annihilation that are moving in opposite directions, the  linear polarizations of the photons must be orthogonal to each other. In today's terms, these two photons are quantum entangled.

Wheeler considered that each photon enters a crystal separately, and is scattered by electrons, thus changes the direction of motion, and is then detected separately. There are various possibilities for the direction in which each photon is scattered, and Wheeler proposed to study  the asymmetry between the probabilities of the scattering directions being perpendicular and  being parallel (defined as the difference between the two divided by the sum of the two) if the scattering angles of the two photons are equal (the scattering angle is the deviation from the incoming direction). The asymmetry is related both to the scattering angle and to the difference between  the azimuthal angles of the two photons. Further careful calculations were made by the group of J. C. Ward and M. Pryce and by the group of H. Snyder, S. Pasternack and J. Hornbostel.   It was calculated that the asymmetry reaches the maximum   $2.85$ when the scattering angle is $85^{\circ}$.    

Before Wu, experiments by others had been done, but none were good. In 1949, Wu and Shaknov gave the final word (published on the first day of 1950). Using a photon detector  10 times more sensitive than the others, they did  an experiment that well confirmed the theory~\cite{12}.  In considering the angular spread around the optimal angle, the maximum of asymmetry is $2.00$.  The Wu-Shaknov experimental value  was $2.4 \pm 0.08$.

In 1935, Einstein, Podolsky and Rosen proposed that quantum mechanics does not satisfy the so-called local realism, which  is analogous to classical probability theory, where it is assumed that there is an objective reality behind the probabilities, even though we do not know the values of certain variables. Could quantum mechanics be reduced to such a situation? Einstein said no. But he felt that local realism always holds and cannot be violated, so he said that there were problems with quantum mechanics and that it was not yet a complete theory. Their argument used quantum entangled states, although they didn't use the term quantum entanglement. After their paper was published, Schr\"{o}dinger coined the term quantum entanglement.

In 1957, the year of the discovery of the parity nonconservation, David Bohm and his student Yakir Aharonov pointed out~\cite{18}  that the experiment of Wu and Shaknov had produced an entangled state of photon polarization, and that Bohm and Aharonov had used the term ``correlation'' to stand for quantum entanglement and demonstrated that the experimental results of  Wu and Shaknov would not have been possible if the photons had not been entangled (or ``correlated'' in their term). This is why the work of Wu is so relevant to quantum entanglement. The experiment of  Wu and Shaknov was the first experimental realization of a spatially well-separated quantum entangled state.

From today's viewpoint of quantum entanglement, Wu-Shaknov result  showed that the two photons were indeed quantum entangled, that is, instead of a product of separate quantum states of individual photons, and hence independent of each other, the quantum state  of the two photons is  a single
entity $\frac{1}{\sqrt{2}}(|h\rangle|v\rangle - |v\rangle|h\rangle)$, where $ |h\rangle$ and $|v\rangle $ are horizontal and vertical polarized states of each photon, respectively. 

In the 1964, John Bell proposed an inequality that is always satisfied by local realism (hidden variable theories), which came to be known as Bell's inequality, and there are some entangled states in quantum mechanics that violate Bell's inequality, so if it is experimentally shown that Bell's inequality is indeed violated, in agreement with quantum mechanics, then it implies that there is something wrong with local realism.

In order to test Bell's inequality, an experimental setup needs to satisfy certain conditions, where the polarizations measured on both sides are neither parallel nor perpendicular. However, the  polarizations measured in the experimental setup of Wu and Shaknov could only be either parallel or perpendicular to each other, so  could not be used to test Bell's  inequality.

In 1975,   Wu decided to return to this field and with two students, L. R. Kasday, J. D. Ullman, and measured coincidence of two photons in a wide range of polar and azimuthal angles, which can be neither parallel nor perpendicular to each other, and obtained the experimental result~\cite{19}. The experimental result could be regarded as a demonstration of quantum entanglement, being consistent with the entangled state and contradict the non-entangled state.  But can this result be used to test Bell's inequality? In fact, it cannot be used to genuinely test the inequality  because here the detection of the photons is through Compton scattering, and after Compton scattering, the direction of flight of each photon is not locked to its polarization direction, i.e. the flight direction  does not in one to one  correspondence  to the direction of polarization, but there is a probability distribution. Indeed, calculation based on local realism can also give the results of Compton scattering. Nevertheless,  if two additional assumptions are made:  (1) the polarization can be measured perfectly,  (2)  the quantum mechanical formulation of Compton scattering is correct, then the experimental results are consistent with quantum mechanics and inconsistent with Bell's inequality.

Later studies of entangled photons were done with low-energy photons that could be measured directly with a polarizer~\footnote{In a few months after the present speech, 2022 Nobel Prize in Physics was awarded to  Alain Aspect, John Clauser and Anton Zeilinger, “for experiments with entangled photons, establishing the violation of Bell inequalities and pioneering quantum information science”. I further discussed the historic role of Wu-Shaknov experiment in  Y. Shi, {\em The road of quantum entanglement: from Einstein to 2022 Nobel Prize in Physics}, Chinese Journal of Nature  {\bf 44} (6),  455-465 (2022), available at
\href{https://www.nature.shu.edu.cn/CN/10.3969/j.issn.0253-9608.2022.06.005}
{https://www.nature.shu.edu.cn/CN/10.3969/j.issn.0253-9608.2022.06.005}, 
  and Y. Shi, {\em Historic origin of quantum entanglement in particle physics: C. S. Wu, T. D. Lee, C. N. Yang and Other Predecessors}, Micius Forum, March 17, 2023, available at   \href{https://mp.weixin.qq.com/s/gs3UxMjvXv1ert1kPu8npg}
  {https://mp.weixin.qq.com/s/gs3UxMjvXv1ert1kPu8npg}; Y. Shi, {\em Historic origin of quantum entanglement in particle physics}, Progress in Physics {\bf 43} (3), 57-67 (2023).  available at \href{https://pip.nju.edu.cn/CN/10.13725/j.cnki.pip.2023.03.001}
  {https://pip.nju.edu.cn/CN/10.13725/j.cnki.pip.2023.03.001}. }.  High-energy photons can destroy the polarizers. Contemporary quantum information science is booming and low energy photons are among the major agents of it.

\section{Details on parity nonconservation}

Now we look into the details of the discovery of parity nonconservation in weak interactions.

In 1956, one of the mysteries at the forefront of particle physics was the $\theta-\tau$ puzzle.  The $\theta$ and $\tau$ are two strange particles with the same mass and same  lifetime, so appear to be a same particle, but they decay differently, into $2$ and $3$  pions respectively. In particle physics, the parity of each particle is  $1$ or $-1$, and the total parity of a group of particles is the product of the parities of all these particles.  Now, the parity of each pion  is $-1$, so if one assumes that the total parity is conserved during the decay process, one can deduce that the parities of $\theta$  and $\tau$ are not the same, being $1$ and $-1$ respectively.  So there are two possibilities. One possibility is that the two particles are the same particle, and parity is not conserved. The other possibility is that the parity  is  conserved but the two particles are different, however it is then hard to understand why they have exactly the same mass and lifetime. At that time, there was a confusion of ideas in the physics community, and whether or not parity is  conserved was discussed without distinguishing  different kinds of interactions. 

The ideas of Chen Ning Yang (then based in Institute for Advanced Study at Princeton, but was at Brookhaven Laboratory for an extended visit)  and Tsun-Dao Lee (at Columbia University) were as follows~\cite{20,21}.  Regarding parity, one should distinguish between two kinds of processes, one is the kind of   processes of particle production, which is dominated by strong interactions and conserves parity, and the other is the kind of processes of particle decays, which is dominated by weak interactions, in which whether the parity is conserved  or not needs to be tested; thus  the $\theta-\tau$  puzzle was extended to a general problem  of  weak interactions, that is, a problem in  the whole area of weak interactions,  allowing  the possibility that parity is conserved in strong and electromagnetic interactions, but may  not be conserved in weak interactions. Therefore knowledge on this problem can be gained by studying other weak interaction processes. If the weak interactions violate parity conservation, then $\theta$ and $\tau$  could be the same particle, decaying into different final states with different parities,  and the $\theta-\tau$ puzzle would be resolved. They sorted out the problem.

Yang and Lee turned to other systems dominated by weak interactions. The most studied process of  weak interactions is $\beta$ decay, so they studied whether or not parity is conserved in $\beta$ decay. This was a major strategic shift that played a key role in the final victory.

First, they naturally examined whether previous experiments on $\beta$ decay had already decided  whether or not parity is conserved. They did  specific  calculations. In weak interactions, in addition to parity-conserving terms, they took into account  additional terms that did not conserve parity, and then calculated the experimentally  observable quantities, especially the distribution of electrons produced from  $\beta$ decay.

Tsung-Dao Lee  and Chien-Shiung Wu  were both in the physics department at Columbia University, so in May, Lee went to talk to Wu, an expert on  $\beta$ decay, who lent Lee  the anthology on $\beta$ decay edited by K. Siegbahn.   Yang had worked together with Jayme Tiomno on $\beta$  decay in 1950, and the experience and results came in handy. In fact, this Lee-Yang parity paper  cited the Yang-Tiomno paper.

The calculations by Yang and Lee showed that whether or not parity is conserved does not affect the results of previous experiments, which thereby  cannot be used as evidence of parity conservation (an exception was unearthed later in 1959). They later realized that the reason was that these previous experiments measured scalars, which always remained constant under parity transformation, i.e. spatial inversion or reflection, and did not involve pseudoscalars, which reverse the signs under parity transformation, i.e. spatial inversion or reflection,  so the previous experiments,  had not really  tested  parity conservation; if the pseudoscalars were measured, whether or not they remained unchanged under spatial inversion or reflection depended on whether or not parity is  conserved, and this was what   needed to be experimentally test.

They noted that  one such  pseudoscalar is what is now called helicity, which is the average of the component of some momentum in some spin direction, or vice versa. If parity is conserved, then the helicity is 0; if parity is not conserved, then the helicity is nonzero. They realized that a simple way of verifying this is to measure, for $\beta$ decay of a nucleus with a given spin (called a polarized nucleus), whether it emits as many electrons upwards  as it does downwards.

In Brookhaven Laboratory, Maurice Goldhaber  told Yang that nuclei had already been polarized by low temperature techniques. In Columbia, Lee asked Chien-Shiung Wu whether pseudoscalar quantities like helicity had been measured. Wu said no,  only scalar quantities had been measured, and asked whether anyone had any idea on it. Lee mentioned polarized nuclei from reactors. Wu  had great misgiving about this approach, and suggested to use the  cobalt 60 nuclei polarized by the demagnetization method, which is just the low temperature technique.  Wu later recalled that she made this suggestion also because she had been following this technique for several years, and the spin nature of cobalt 60 does not lead to a reduction of possible asymmetries in decay~\cite{22}.

So polarized cobalt 60 became an experimental  candidate that was  discussed in Lee-Yang  paper. They also discussed mesons and hyperons as  experimental candidates. Firstly, parity nonconservation would lead to an electric dipole moment in a $\Lambda$ hyperon. Second, consider a pion impinging on a proton, producing a $\Lambda$  hyperon,which then decays  into a pion  and a proton, and using the momenta of the three particles, the incident pion, the $\Lambda$  hyperon, and the product pion, to form a vector triple product, which is a pseudoscalar. By measuring whether or not the range of its values  is symmetric, one can test for  parity conservation. Thirdly, consider that a pion decays into  a muon, which further decays into an electron, and if parity is not conserved, the spin of the muon is mostly along the direction of motion,  resembling a polarized nucleus, and the distribution of electrons from the decay of muon is not symmetric.

Immediately after  Wu suggested the use of polarized cobalt 60 to  Lee, it occurred to Wu that this experiment was a golden opportunity for a $\beta$ decay experimentalist and should not be missed, and that even if the experiment would  demonstrate parity conservation, it would still be meaningful in  giving  an upper bound of parity violation and thereby stopping conjectures about it. She also felt that she had to do the experiment immediately before the rest of the physics community realised the importance. For this reason she abandoned her original travel plan to attend an international conference in Geneva and then return to the Far East, even though it would have been her first trip back after 20 years away from China.

Thus Wu began to prepare for the experiment. Firstly, in the new version of the nuclear data, the spin of cobalt 60 was changed, meaning that it was not as suitable as she had originally considered for testing parity conservation. Through experimentation, Wu and her assistant determined that the new value  of the spin was incorrect and that the original one was correct. This also manifested  the scientific spirit  of Wu, who did not follow the crowd. She also learnt low temperature techniques from the low temperature group in her department.

There were two difficulties in this experiment. One was to place the electron detector in a cryostat at the liquid nitrogen temperature and have it work, and the other was to place the sample of cobalt 60 in a surface layer that would remain polarized for a sufficient period of time. The principle of polarization is that a weak magnetic field at low temperatures causes the electrons in a paramagnetic crystal to align themselves in a certain direction, which in turn, through hyperfine interactions, causes the spin of the nucleus to align in  that direction.

On 4 June 1956,  Wu telephoned Ernest Ambler of the National Bureau of Standards in Washington, D.C., inviting him to collaborate.  Amber enthusiastically accepted.  
The  polarization of cobalt 60 had been achieved by the low-temperature physics group led by Nicholas Kurti of Oxford University a few years earlier, and the first author of the paper had been none other than Ambler, who had been Kurti's graduate student.

In June and July,  Wu tested the $\beta$ detector, considering many technical issues.
The theoretical paper by Tsung-Dao Lee and Chen Ning Yang was submitted on 22 June. On 24 July, Wu called Ambler again, who sent a rough drawing of the cryostat a few days later, and then went on holiday for two weeks, according to Wu's reminiscence later~\cite{22}.

In August, Wu studied the effect of magnetic fields on $\beta$ counting, and the scattering of $\beta$  particles by  cerium magnesium nitrate (CMN) crystals.

In mid-September, Wu went to the National Bureau of Standards in Washington, D.C., to meet with Ambler for the first time, and was joined in the collaboration by R. P. Hudson, R. W. Hayward, and D. D. Hoppes; Hudson had also been a student of Kurti. On her third visit to the Bureau of Standards,  Wu brought two CMN crystals containing cobalt 60. A large number of CMN crystals were later grown at Columbia University by Wu and her assistants.

In mid-December, the collaborative group saw for the first time a very large asymmetric signal, meaning that the number of electrons emitted upwards and downwards were very different.

But Wu felt that a systematic examination  was needed to rule out other factors before making an announcement to the public. She later recalled~\cite{22} that one Thursday, while walking past Lee's office in the Pupin Building, the physics Building of Columbia, she encountered Lee and Yang,  and was asked about the experiment.  Wu   told them  that the effect was large and reproducible, but explained that it was only a preliminary result. A week later, the collaborative group began to rule out other possible factors.

Wu also recalled~\cite{22} that on Christmas Eve, in heavy snow, she arrived in New York by train from Washington, D.C., and called Lee  to tell that the asymmetry parameter had almost reached $-1$, meaning that the vast majority of electrons were moving in the direction opposite to the direction of the cobalt 60 polarization.

This result showed the correctness of the two-component theory of neutrino, that is, neutrinos are always left-handed while antineutrinos are always right-handed, which had been worked out by Lee  and Yang  under the premise of parity nonconservation   in the summer, but had not been submitted  because it had not yet been known experimentally whether parity is indeed nonconserved in weak interactions.  Now they submitted this paper, which was received on 10 January~\cite{23}. On the other hand, according to an analysis of the three discrete symmetries (parity,  time reversal, and charge conjugation, i.e., positive-negative particles) just completed by Lee, Reinhard Oehme and Yang, the parity asymmetry is so large  as  found by Wu experiments that the  charge conjugation symmetry should be  violated as well.  The theoretical paper by the three of them was received on 7 January~\cite{24}. Thus, the Wu experiment not only discovered that parity is indeed not conserved, but also   promoted the theoretical work of Lee and Yang.

As mentioned earlier, Yang and Lee's paper also suggested an experiment on the    decay from  pion  to muon  and then to electron $(\pi-\mu-e)$, where the parity nonconservation  leads to an asymmetric distribution of electrons. The Nevis laboratory  affiliated with Columbia University regularly produced this process. At Brookhaven Laboratory, when Yang met Leon M. Lederman from the Nevis Lab, Yang suggested that Lederman to  use  the facilities at the latter's hand  to test  parity conservation,  Lederman joked that he would   do it if he had a smart graduate student as a slave~\cite{20}.

Rumors  about Wu experiment had spread during the Christmas holidays. On 4 January 1957, at a routine Friday Chinese luncheon of Columbia physics department, Lee commented  that Wu experiment had a large effect. Lederman attended this luncheon. That evening, Richard L. Garwin, Lederman and Marcel Weinrich  began the $\pi-\mu-e$ experiment. The experiment was very simple, and the result  was  very clear. At 6:00 a.m. on the 8th, Lederman called Lee: ``the law of parity is  dead.''~\cite{25}.

Now a crisis arose; Wu experiment promoted  Lee-Yang  theory and triggered the $\pi-\mu-e$ experiment, but the final confirmation of the conclusions of the Wu experiment itself had not yet been completed. The 2nd to the 9th of January was the most stressful period for Wu's collaborating group.  

Now the Garwin-Lederman-Weinrich experiment had been done, and although Wu was not very happy, her  group remained relentless until other factors were ruled out and the results were fully confirmed. Wu recalled that at 2:00 a.m. on 9,  the group celebrated their overthrow of the law of parity~\cite{22}.

On the morning of the 13th, in room 831 of the Physics Building at Columbia University, Wu, Garwin, Lederman, Weinrich, Lee  and Yang held a discussion meeting.

On 15 January, a press conference was held at Columbia University, hosted by Isidor   Rabi, a senior member of the Physics Department, and attended by all the members of the two experimental groups and Lee. The papers of the two groups were received by Physical Review  on the same day, and were published back to back on 15 February, with Wu's  in the lead. The following day the New York Times headlined.

Parity  nonconservation in weak interactions was discovered by Wu's group.  The Garwin-Lederman-Weinrich experiment was done only after the results of Wu were known. In addition, Valentine Telegdi of the University of Chicago had been working independently on the $\pi-\mu-e$ experiment since the summer, but proceeded very slowly, and after learning about the news about the Wu experiment and the Garwin-Lederman-Weinrich experiment, rushed to submit a manuscript, which turned out to be problematic and was published one issue late.

Yang and Lee  found, after specific calculations, that there was no previous experimental proof of parity conservation   in weak interactions, and pointed out several types of critical experiments   to test whether parity  is conserved in weak interactions. However, they did not say that parity must be conserved or not, and in fact they also  proposed the parity doublet state in the framework of parity  conservation to solve the $\theta-\tau$ puzzle. Whether or not parity  is conserved requires an experimental ruling.

Parity conservation has long had an intuitive appeal and has been regarded as natural, sacred, and very useful, especially in nuclear physics. As a result, the parity paper of Yang and Lee was universally despised, disagreed with and even ridiculed~\cite{26}.

But it took great courage for Chien-Shiung Wu to decide to do a $\beta$ decay experiment on cobalt 60. In Yang's words, the insight of Wu was unique. If the Lee-Yang paper had given a clear theoretical prediction of whether or not parity is conserved in weak interactions, there would have been many people rushing to do such  experiments instead.

Wu's initial decision to go ahead with the experiment and her insistence on continuing to verify the experimental result after the result of the Garwain-Lederman-Weinrich experiment all demonstrated the greatness of Wu.

The Wu experiment proved that parity is indeed non-conserved in $\beta$ decay, causing a huge shock to the entire physics community, and became one of the most important experiments in physics in the 20th century.

James W. Cronin, who won the 1980 Nobel Prize for his discovery of charge-conjugation-parity (CP) nonconservation,  once said, ``The great discovery of Chien-Shiung Wu started the golden age of particle physics.'' It is a mistake and a pity that the Nobel Prize was never awarded to  Wu. Of course, the origin of this great discovery was the pioneering theoretical work of Yang and Lee.

When I was a student at Nanjing University, I had the privilege of listening to a lecture given by   Wu as an alumnus of the University, and I still remember the scene when she said the English word ball,  probably  talking about the cobalt 60 nucleus.

\section{Chien-Shiung Wu: A Great Experimental Physicist} 

In conclusion, Chien-Shiung Wu was a great experimental physicist. 

We can summarize some of the qualities of  Chien-Shiung Wu: intelligent, calm, diligent, hard-working, focused, undaunted by difficulties, and dedicated to science. She has done what her father said when she was a child: ``Fear no difficulties, work hard, and keep moving forward''.

As Chen Ning Yang  said, ``The research works of Ms Chien-Shiung Wu were well known for their  precision and accuracy''. This is related to characteristic of many  Chinese women, but it is also a manifestation  of her scientific spirit.    I hope this   encourage more women to devote themselves to science.

Wu's  research areas and achievements were closely related to her years of experience  and expertise.  She attached importance to the development of experimental techniques and instruments, as well as to physical significance, following  important theories and verifying them with experiments. She has been working in related fields, especially $\beta$ decay, since her student days.

Her experimental work had significant theoretical implications, such as verifying Fermi's theory of $\beta$ decay, the prediction of quantum electrodynamics on photon coincidence and quantum entanglement, parity  nonconservation, conservation of vector current, and double $\beta$ decay, indicating  that she kept track of the cutting-edge theories of the time.

Whether she had to falsify or confirm an important theory, she went for it, believing that even if it seemed obvious to common sense, if experimental evidence was lacking, it was worth doing.

I think the most valuable thing of all: putting scientific rigour above competition and honour. This is a great example of her scientific spirit. She was competitive, but worked under scientific rigour even when priority and credit were at stake. She insisted on double-checking her own results when priority and credit were at risk, and when others had already obtained very clear experimental results. This is very valuable. Indeed, the Wu experiment on parity nonconservation, while respected for publication, did suffer from a loss of priority. This was one of the factors in her losing the Nobel Prize. If the Garwin-Lederman-Weinrich experiment had not existed, there would have been a larger possibility that she would have shared the Nobel Prize with Yang and Lee, and the other factors  for not winning the Nobel Prize would have been difficult to work.

Chen Ning Yang's comment on Chien-Shiung Wu  quoted above  was from a eulogy written by  Yang in April 1997   following  Wu's death in February that year (translated from Chinese by Yang himself): ``The research works of Ms Chien-Shiung Wu were well known for their  precision and accuracy. But her great success was down to  another more fundamental reason:  in 1956, people did not want  to do experiments to test parity conservation. Why was she willing to do this such a difficult experiment? Because she had the penetrating perception that even if parity conservation is not overthrown, this  fundamental law of nature must be  tested. Here resides  her greatness.''

Later, Yang also wrote: ``I once said that there are three necessary conditions for success in scientific research: Perception,Persistence, and Power. Chien-Shiung Wu satisfied all these three conditions. Her experiment on parity nonconservation met with many  difficulties. The 9th chapter of the Biography of Chien-Shiung Wu   vividly describes how she persevered and overcame all kinds of difficulties with her Power. The most important thing was her vision: at that time, many other first-rate physicists thought that such a difficult experiment was not worth doing because it was just another proof that the parity is indeed conserved. But she had the unique insight that the conservation or not  of parity  in weak interactions had not been studied in the past, so it was a worthwhile experiment regardless of the results. This is where her vision was exceptional.''~\cite{27}. 

Tsung-Dao Lee said: ``She was one of the most outstanding physicists of the 20th century, who made great achievements in experimental physics research and played an extremely important role in advancing the development of contemporary physics.''~\cite{9}

Lee also said: ``When Madam Curie passed away, Albert Einstein once wrote, `At a time when a towering personality has come to the end of her life, let us not merely rest content with recalling what she has given to mankind in the fruits of her work.  It is the moral qualities of its leading personalities that are perhaps of even greater significance for a generation and for the course of history than purely intellectual accomplishments $\cdots$ Her strength,her purity of will $\cdots$ her objectivity, her incorruptible judgment, all these were of a kind seldom found joined in a single individual $\cdots$ Once she had recognized a certain way as the right one, she pursued it without compromise and with  extreme tenacity.' I think that in remembering Chien-Shiung Wu,  it is only fitting that we  apply to her the words of Einstein praising Madame Curie.''~\cite{9}

The scientific life of Chien-Shiung Wu is situated in the context of 20th century history of  China  and the world.  Segr\`{e},  the mentor of   Wu,  wrote a passage about Wu, Yang and Lee,  which was later quoted by Yang  in his speeches: ``This trio of Chinese physicists shows what China’s future contribution to physics could be if that great country overcomes the period of revolutionary convulsions and resumes its historic role as one of the leaders of civilization, as witnessed by the early European travelers, to their astonishment.''~\cite{28} 

\section*{Acknowledgment}

Throughout  the years, I have made numerous discussions with Prof. Chen Ning Yang regarding topics covered in this speech. Prof. Yang  encouraged me to write contributions of Chien-Shiung Wu, including her early work on quantum entanglement, leading to my  English note written in early 2022~\cite{mpla} and the present speech. This work is supported by National Nature Science Foundation (Grant No. T2241005 and No. 12447214).


\begin{thebibliography}{99} 

\bibitem{1} H. Yang, Madame Curie's Chinese disciple (in Chinese), Modern Physics  Knowledge {\bf 19}(6), 67-68 (2007).

\bibitem{2} T. C. Chiang,  C.S. Wu: The First Lady of Physical Science (in Chinese), Fudan University Press, Shanghai 1997.  English Translation: T. C. Chiang, tranlated by T. F. Wong,   Madame Wu Chien-Shiung: The First Lady of Physics Research, World Scientific, Singapore (2014).

\bibitem{3}  C. S. Wu, The continuous x-rays excited by the $\beta$-particles of P-15 (32), Phys. Rev. {\bf 59} (6),481-488 (1941).

\bibitem{4}  C. S. Wu and E. Segrè, Some fission products of uranium, Phys. Rev. {\bf 57} (6), 552-552 (1940).

\bibitem{5} C. S. Wu, Identification of two radioactive xenons from uranium fission, Phys. Rev. {\bf 58} (10),  926-926 (1940).

\bibitem{6} C. S. Wu and E. S\`{e}gre, Radioactive Xenons,  Phys. Rev.  {\bf 67} (5-6), 142-149 (1945)

\bibitem{7} C. S. Wu and G. Friedlander,  Radioactive isotopes of mercury,   Phys. Rev.  {\bf 60}(10), 747-748 (1941).  

\bibitem{8}  G. Friedlander and C. S. Wu,  Radioactive isotopes of mercury,   Phys. Rev.  {\bf 63}(7/8), 0227-0234 (1943).   

\bibitem{9}  Tsung-Dao Lee, Chien-Shiung Wu and the experiment on parity nonconservation,  Science (Chinese) {\bf 49} (5), 3-10 (1998).

\bibitem{10} C. S. Wu and R. D. Albert, The $\beta$-ray spectra of $Cu^{64}$ and ratio of $N_+/N_-$,  Phys. Rev. {\bf 75} (7) , 1107-1108 (1949).   

\bibitem{11}  C. N. Yang, C. S. Wu's Contributions: a Retrospective in 2015, Int. J. Mod. Phys. A {\bf 30} (20), 1530050 (2015).

\bibitem{12} C. S. Wu and I. Shaknov,  The angular correlation of scattered annihilation, Phys. Rev. {\bf 77}(1), 136-136 (1950).  

\bibitem{13} T. D. Lee and C. N. Yang, Question of parity conservation in weak interactions. Phys. Rev. 104 (1), 254-258 (1956).

\bibitem{14}  C. S. Wu, E. Ambler, R. W. Haywood, D. D. Hoppes and R. P. Hudson, Experimental Test of Parity Conservation
in Beta Decay, Phys. Rev. {\bf 105}(4), 1413-1415 (1957).  

\bibitem{15} Y. K. Lee, L. W. Mo and C. S. Wu,  Experimental test of conserved vector current theory on $\beta$ spectra of B12 and N12, Phys. Rev. Lett. {\bf 10} (6), 253 (1963).  

\bibitem{16}  C. S. Wu, The Universal Fermi Interaction and the Conserved  Vector Current in Beta Decay, Rev. Mod. Phys. {\bf 36}, 618-632 (1964). 

\bibitem{17}  Chien-Shiung Wu, The state of US physics-1976, Physics Today {\bf 29}, 23-30 (1976).

\bibitem{18}  D. Bohm, Y. Aharonov, Discussion of Experimental Proof for the Paradox of Einstein, Rosen, and Podolsky, Phys. Rev. {\bf 108}, 1070 (1957).

\bibitem{19} L. R. Kasday, J. D. Ullman and C. S. Wu,  Angular correlation of compton-scattered annihilation photons and hidden variables,  Nuovo Cimento B {\bf 25}(2), 633-661 (1975).  

\bibitem{20}  C. N. Yang, Selected Papers 1945-1980 With Commentary (W. H. Freeman and Company Publishers, 1983).

\bibitem{21}  G. Feinberg (Ed.), T.D. Lee: Selected Papers. Volume 3, Birkhauser, Boston, 1986.
    
\bibitem{22}  B. Maglich (Ed.), Adventures in experimental Physics, $\gamma$  Volume, World Science Education (January 1, 1973).  
    
\bibitem{23}  T. D. Lee and C. N. Yang, Parity Nonconservation and a Two-Component Theory of the Neutrino, Phys. Rev. {\bf 105}, 1671 (1957).
    
\bibitem{24} T. D. Lee, R. Oehme and C. N. Yang, Remarks on possible
noninvariance under time reversal and charge conjugation, Phys. Rev. {\bf 106} 
(2), 340-345 (1957).
\bibitem{25}  L. Lederman and D. Teresi, The God Particle, (Houghton Mifflin, New York, 1993).
\bibitem{26} Y. Shi, 13 important scientific contributions of Chen Ning Yang,  Physics (Chinese), {\bf 43} (1), 57-62  (2014); 
Yu Shi,  Beauty and physics: 13 important contributions of Chen Ning Yang,  Int. J.  Mod. Phys.  A  {\bf 29}, 1475001 (2014);
Yu Shi, Brief Overview of C. N. Yang's 13 Important Contributions to Physics,  Int. J.  Mod. Phys.  A {\bf 30}, 36 (2015). 
\bibitem{27} C. N. Yang’s Dawning Volume (Chinese), ed. F. Weng  (World
Scientific, Singapore, 2008). 
\bibitem{28} E. Segr\`{e}, From X-rays to Quarks: Modern Physicists and Their
Discoveries, (Dover Publication, Mineola, 2007).
\bibitem{mpla}  Y. Shi, Chien-Shiung Wu as the experimental pioneer
in quantum entanglement: A 2022 note, Mod. Phys. Lett. A
{\bf 40},  2530001 (2025);  also available at \href{https://arxiv.org/abs/2502.06458}{https://arxiv.org/abs/2502.06458}
\end{thebibliography}
\end{document}